%% file: x4140_pub.tex
\RequirePackage{lineno}

\documentclass[aps,prd,twocolumn,showpacs,superscriptaddress,groupedaddress]{revtex4} 

\usepackage{graphicx}
\usepackage{dcolumn}
\usepackage{bm}

\begin{document}

\hspace{5.2in} \mbox{FERMILAB-PUB-15-373-E}

\title{Inclusive production of the {\boldmath $X(4140)$} state in {\boldmath $p \overline p $} collisions  at D0}

\input author_list.tex

\date{August 31 2015}
           
\begin{abstract}
We present a study of the inclusive production of the  $X(4140)$ with the decay to the
$J/\psi \phi$ final state
in hadronic collisions. Based on  $10.4~\rm{fb^{-1}}$ of $p \overline p $ collision
data  collected by the  D0 experiment at  the Fermilab Tevatron collider, we report the first evidence
 for the prompt production of  $X(4140)$  and  find the fraction of $X(4140)$ events originating 
from $b$ hadrons  to be $f_b=0.39\pm 0.07  {\rm \thinspace (stat)} \pm 0.10 {\rm \thinspace (syst)} $.
 The ratio of the non-prompt  $X(4140)$  production rate to the 
 $B_s^0$ yield in the same channel is
 $R=0.19 \pm 0.05  {\rm \thinspace (stat)} \pm 0.07  {\rm \thinspace (syst)}$.
 The  values of the mass $M=4152.5 \pm 1.7 (\rm {stat}) ^{+6.2}_{-5.4} {\rm \thinspace (syst)}$~MeV and width
 $\Gamma=16.3 \pm 5.6 {\rm \thinspace (stat)} \pm 11.4 {\rm \thinspace (syst)}$~MeV
are consistent with previous measurements.
 
\end{abstract}


\maketitle


\newpage

The $X(4140)$ state~\cite{pdg2014} was first seen in 2009 as a narrow structure  in the $J/\psi \phi$ system  near threshold.  The CDF Collaboration reported the first 
evidence~\cite{Aaltonen:2009at} for this state (then designated  $Y(4140)$) in the decay
 $B^+ \rightarrow X(4140) K^+ \rightarrow J/\psi \phi
 K^+$ (charge conjugation is implied throughout this paper) and measured the invariant mass $M=4143.0 \pm2.9 (\rm {sta
t}) \pm 1.2 {\rm \thinspace (syst)}$~MeV and width $\Gamma= 11.7^{+8.3}_{-5.0} {\rm \thinspace (stat)} \pm 
3.7 {\rm \thinspace (syst)}$~MeV.  
  The LHCb Collaboration found no evidence for the $X(4140)$ state~\cite{Aaij:2012pz} in a 2.4 standard deviation
 disagreement with the CDF measurement.
However, the presence of $X(4140)$ in $B^+$ decay was later confirmed by the CMS~\cite{cms} and D0~\cite{d0x}
 Collaborations. The BaBar Collaboration searched for resonant production in the $J/\psi \phi$ mass spectrum
in $B^{+,0}$ decays and obtained a significance below 2$\sigma$, but noted that the hypothesis that the events are distributed uniformly on the Dalitz plot gives a poorer description of the data~\cite{babar}. 
The  quantum numbers of the   $X(4140)$ state have not been measured. 
Since both the $J/\psi$ and $\phi$ mesons have $I^GJ^{PC}=0^-1^{--}$, the  state has positive $G$ 
and $C$ parities.

 A meson decaying into a charmed quark pair  might be an excited charmonium state.
  However, the standard nonrelativistic quark model of  a single $c\overline c$ pair
 does not predict a hadronic state at this mass. Also, at  masses above the open-charm threshold of 3740~MeV
 such states  are expected to  decay predominantly to pairs of charmed mesons and to have a much larger
 width than is experimentally observed. It has been suggested that $ X(4140)$ could be a molecular 
structure made of two charmed mesons, e.g. $(D_s,\overline D_s)$.
Other possible states  are hybrids  composed of two quarks and a 
valence gluon ($q\overline qg$), four-quark combinations ($c\overline c s \overline s$), or states with
  higher Fock components~\cite{azimov}. For details see the reviews  in
 Ref.~\cite{Drenska:2010kg} and \cite{kai} and references therein.
The Belle Collaboration found no evidence for  $X(4140)$ in the process
 $\gamma \gamma \rightarrow J/\psi \phi$~\cite{Shen:2009vs}, making its interpretation as a hadronic
 molecule with spin-parity  $J^P =0^+$ or $2^+$ unlikely.  

In addition to $X(4140)$, the CDF Collaboration reported seeing a second enhancement in the same channel, 
located near 4280 MeV.
 A similar structure is seen by the CMS Collaboration~\cite{cms}
at a slightly higher mass of $4316.7 \pm 3.0 {\rm \thinspace (stat)} \pm 7.3 {\rm \thinspace (syst)}$~MeV.
 Belle also reports a new structure
at $M=4350.6^{+4.6}_{-5.1}{\rm \thinspace (stat)}\pm0.7  {\rm \thinspace (syst)}$~MeV.

In this Article we present results of a search for the $X(4140)$ resonance   in the $J/\psi \phi$ system
produced inclusively in  $p \overline p $ collisions, either promptly, by pure QCD,
or through weak decays of $b$ hadrons.  The measured production rates are normalized to
the rate of the process $B_s^0 \rightarrow J/\psi \phi$ measured with  the same dataset.
The data sample corresponds to an integrated luminosity of 10.4~fb$^{-1}$
 collected with the D0 detector  in $p \overline p $ collisions at 1.96 TeV at the Fermilab Tevatron 
collider.


The D0 detector consists of a central tracking system, calorimeters, and
muon detectors~\cite{Abazov2006463}. The central
tracking system comprises  a silicon microstrip tracker (SMT) and a central
fiber tracker (CFT), both located inside a 1.9~T superconducting solenoidal
magnet.  The tracking system is designed to optimize tracking and vertexing
for pseudorapidities $|\eta|<3$,
where  $\eta = -\ln[\tan(\theta/2)]$, and  $\theta$ is the 
polar angle with respect to the proton beam direction.
  The SMT can reconstruct the $p\overline{p}$ interaction vertex (primary vertex) 
for interactions   with at least three tracks with a precision
of 0.004~cm in the plane transverse to the beam direction.
  The muon detector, positioned outside the calorimeter, consists of a central muon system covering the
 pseudorapidity region  $|\eta|<1$ and a forward muon system covering the pseudorapidity region
  $1<|\eta|<2$. 
Both central and forward systems consist of a layer of drift  tubes
and scintillators inside 1.8~T iron toroidal magnets with two similar layers outside the toroids~\cite{muid}.

Events used in this analysis are collected with both single-muon and dimuon triggers.
Muon triggers require a coincidence of signals in trigger elements inside and outside
the toroidal magnets.
Dimuon triggers in the central rapidity region  require at least one muon to penetrate the toroid.
In the forward region, both muons are required to penetrate the toroid.

We study a wide range of the  $J/\psi \phi$ invariant mass,
from threshold  to 5.7~GeV, covering both the $X(4140)$ and the decay $B_s^0 \rightarrow J/\psi \phi$.
Candidate events are required to include a pair of  oppositely charged muons in the invariant mass range $2
.9<M(\mu^+ \mu^-)<3.3$~GeV, consistent with $J/\psi$ decay, accompanied by two additional  particles of
 opposite charge, assumed to be kaons,  with $p_T>0.4$~GeV  and
  $1.011 <M(K^+K^-)<1.030$~GeV.   In the event selection, both muons are required to be
 detected  in the muon chambers inside the toroidal magnet, and at least one of the muons is required to be also 
detected
 outside the iron toroid~\cite{muid}.
Each muon candidate is required to match a track found in the central tracking system, and
 each of the four final-state tracks is required to have at least one SMT hit and at least one CFT hit.
The dimuon invariant mass is constrained  to the world-average   $J/\psi$ mass~\cite{pdg2014},  and 
 the four-track system   is constrained to a common vertex.
To reconstruct the primary vertex, tracks are selected that 
do not originate from the $J/\psi \phi$ candidate, 
and a constraint is applied to the average beam  position in the transverse plane.
We require  $J/\psi \phi$ candidates to have  $5<p_T<20$ GeV and rapidity $|y|<2$.

\begin{figure}[h!tb]
   \includegraphics[width=0.8\columnwidth]{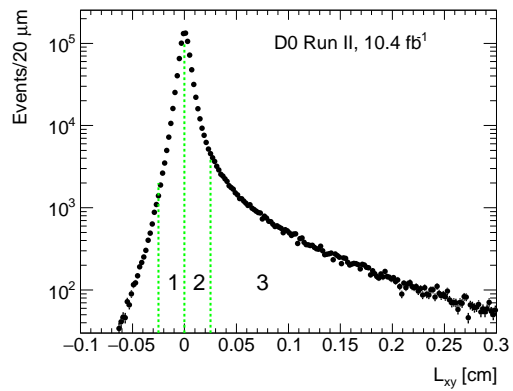}
  \caption{Transverse decay-length distribution of  $J/\psi \phi$ candidates.
The vertical lines define the three regions discussed in the text.
 }
   \label{fig:lxy} 
\end{figure}

We define the signed  decay length of the $J/\psi \phi$ system, $L_{xy}$, 
to be  the  vector pointing
from the primary vertex to the decay vertex, projected onto the direction of the transverse momentum. 
The distribution of  $L_{xy}$  for the selected events  is shown in Fig.~\ref{fig:lxy}.

We focus on two ranges of the  $J/\psi \phi$ invariant mass, $M(J/\psi \phi)<4.36$ GeV
and  $4.8<M(J/\psi \phi)<5.7$ GeV. The low-mass range includes the $X(4140)$ state.
The high-mass range includes the reference decay process,  $B_s^0 \rightarrow J/\psi \phi$.
Background arises primarily from non-resonant pairs in the $\phi$ mass window. At low $L_{xy}$, 
background comes from $J/\psi$ mesons directly produced in  $p\overline p$ collisions combined 
with random particles from the underlying event. At higher values of $L_{xy}$,
background consists of $J/\psi$ mesons paired with  random  products of $b$ hadron decays.  

We divide the data in each mass range into three independent subsamples according to the value of  $L_{xy}$:
(1)  $-0.025\le L_{xy}<0$ cm,  (2)  $0\le L_{xy}\le 0.025$ cm, and (3)  $L_{xy}>0.025$ cm.
Region 1 includes half of the prompt events and almost no $B$-decay events (the fit result shown
in Table~\ref{tab:results}  is $37 \pm 26$ events).
Region 2 includes the remaining half of all prompt events and a fraction of non-prompt events.
The rest of the non-prompt events populate region (3). Given the average resolution of 0.006 cm
in $L_{xy}$, we assume that
the fraction of prompt events in Region 3 is negligible.
 We perform binned maximum likelihood fits to the distributions of the  $J/\psi \phi$ invariant mass for
 events in the six subsamples
 defined above.  In the fits in the $B_s^0$ mass region,  the signal is described by a Gaussian function
 and
background is described by a second-order Chebychev  polynomial. 
We also allow for the presence of the decay $B^0 \rightarrow J/\psi \phi$,
 where we set the mass to the world-average $B^0$ mass, and  we find no evidence of a signal.
  The fit  for Region 3  yields $3166\pm81$   $B_s^0$  events.

In fitting the low mass range,  we assume a  signal described by  an $\cal S$-wave 
 relativistic Breit-Wigner function convolved with a Gaussian resolution
of $\sigma(M)=4$~MeV.  The background is parametrized by the function
 $f(m) \propto m\cdot (m^2/m_{\rm thr}^2 -1)^{c_1}\cdot e^{-m\cdot c_2}$
where $m_{\rm thr}$  is the kinematic threshold, and  $c_1$ and $c_2$ are free parameters. 
For events in the $L_{xy}$ Region 3,    we allow the  signal mass and width parameters
to vary. 
The fit yields $616\pm170$   signal events, a mass of $4152.5\pm1.7$ MeV,
and a width of $16.3\pm 5.6$ MeV.
 The statistical significance of the signal, 
based on the increase of the likelihood with respect to the fit with no signal, $-2\Delta \ln{\cal L}=42.5$
 for 3 degrees of freedom, is 5.9 standard deviations. 
For the fits in $L_{xy}$ Regions 1 and 2 we set the mass and width to the Region 3 values.

\begin{figure*}[htbp]
  \includegraphics[width=0.8\columnwidth]{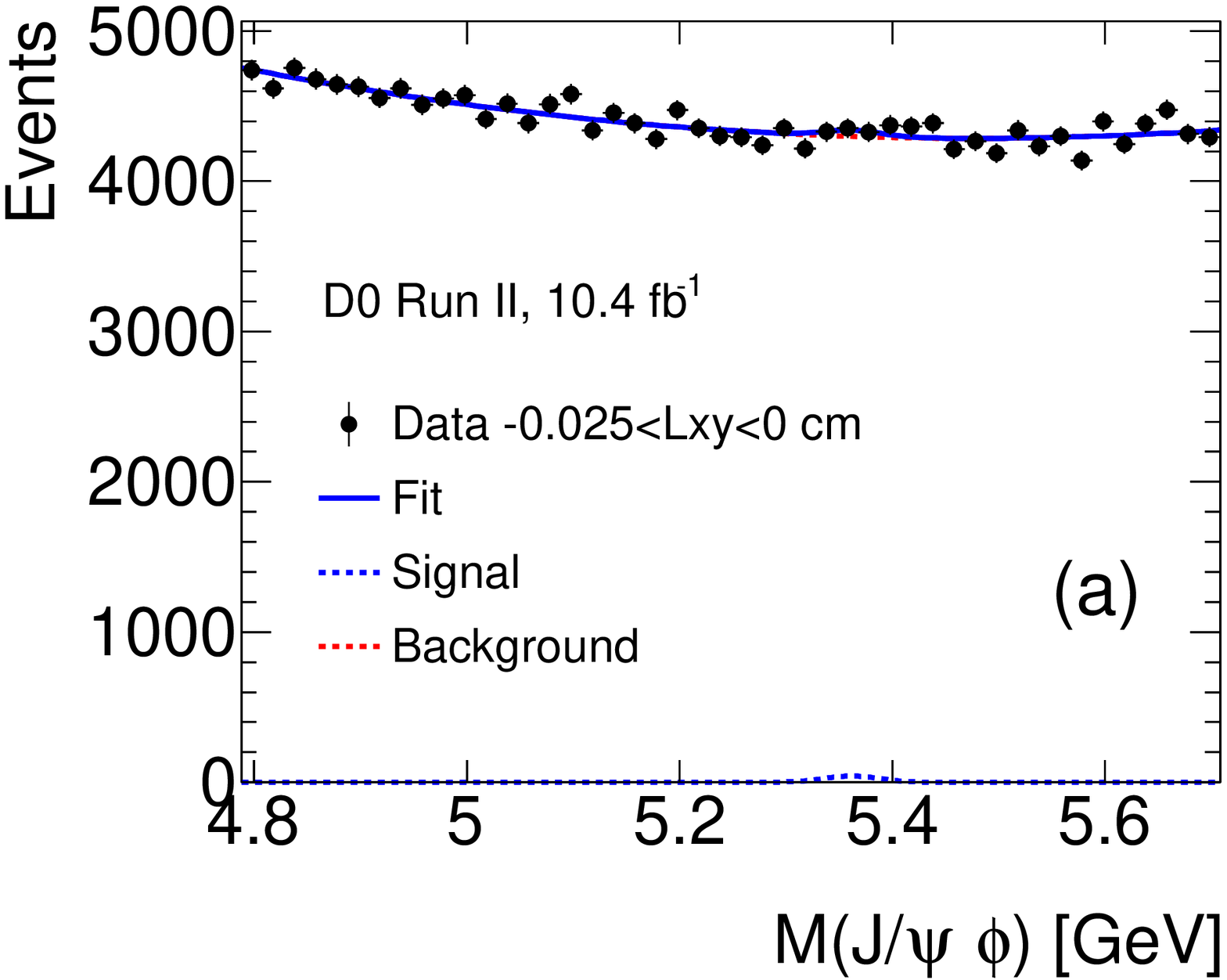}
  \includegraphics[width=0.8\columnwidth]{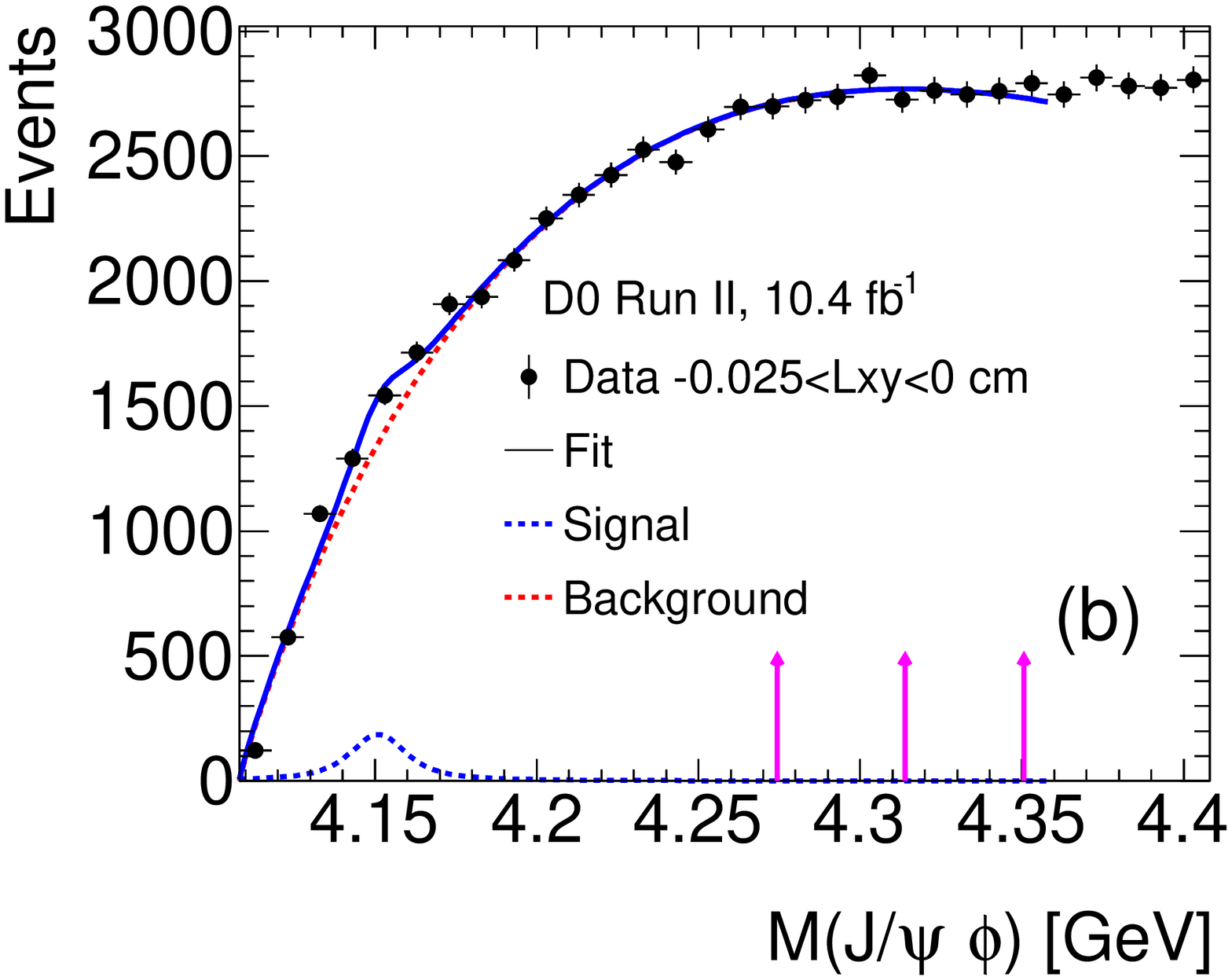} 
  \includegraphics[width=0.8\columnwidth]{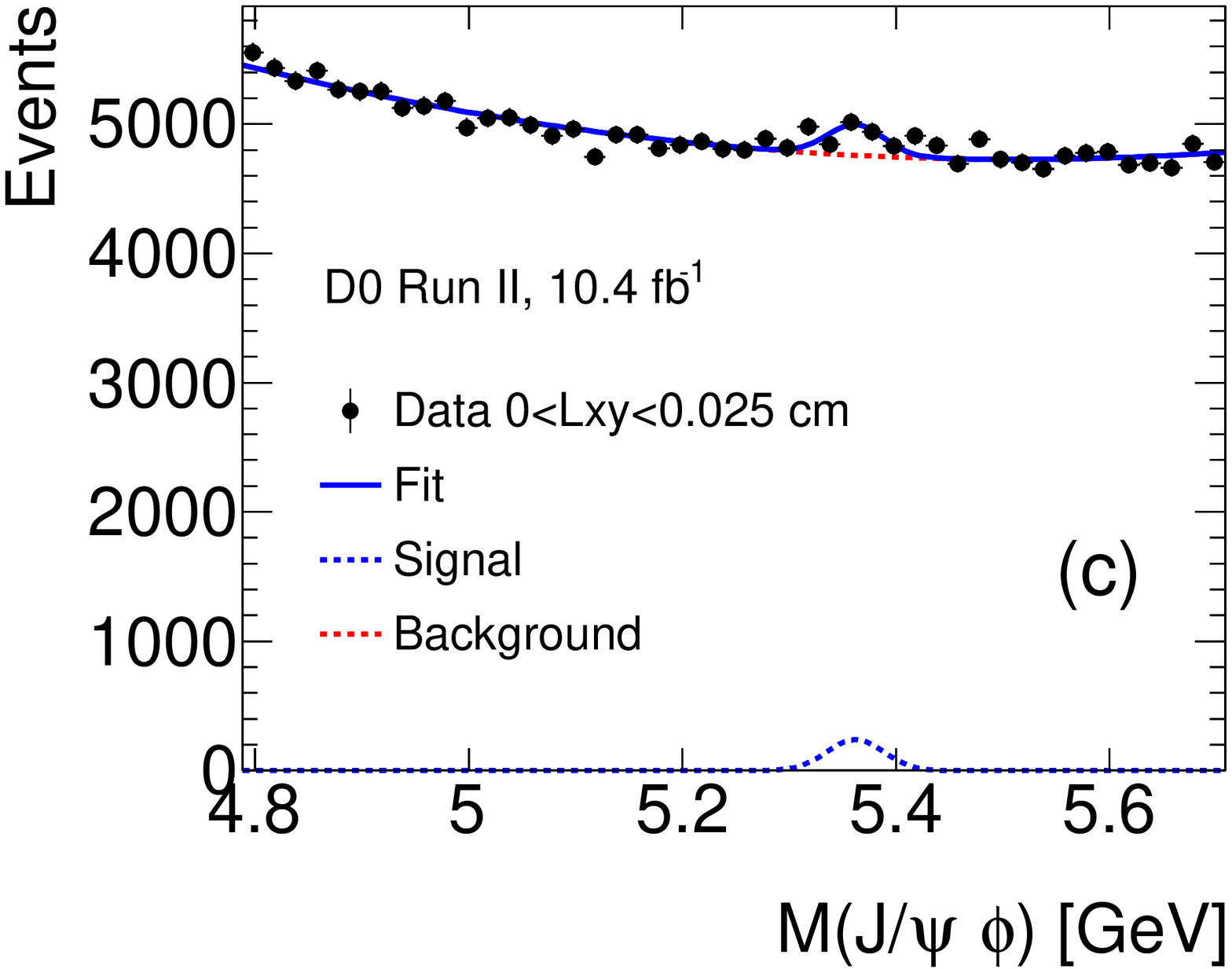}
  \includegraphics[width=0.8\columnwidth]{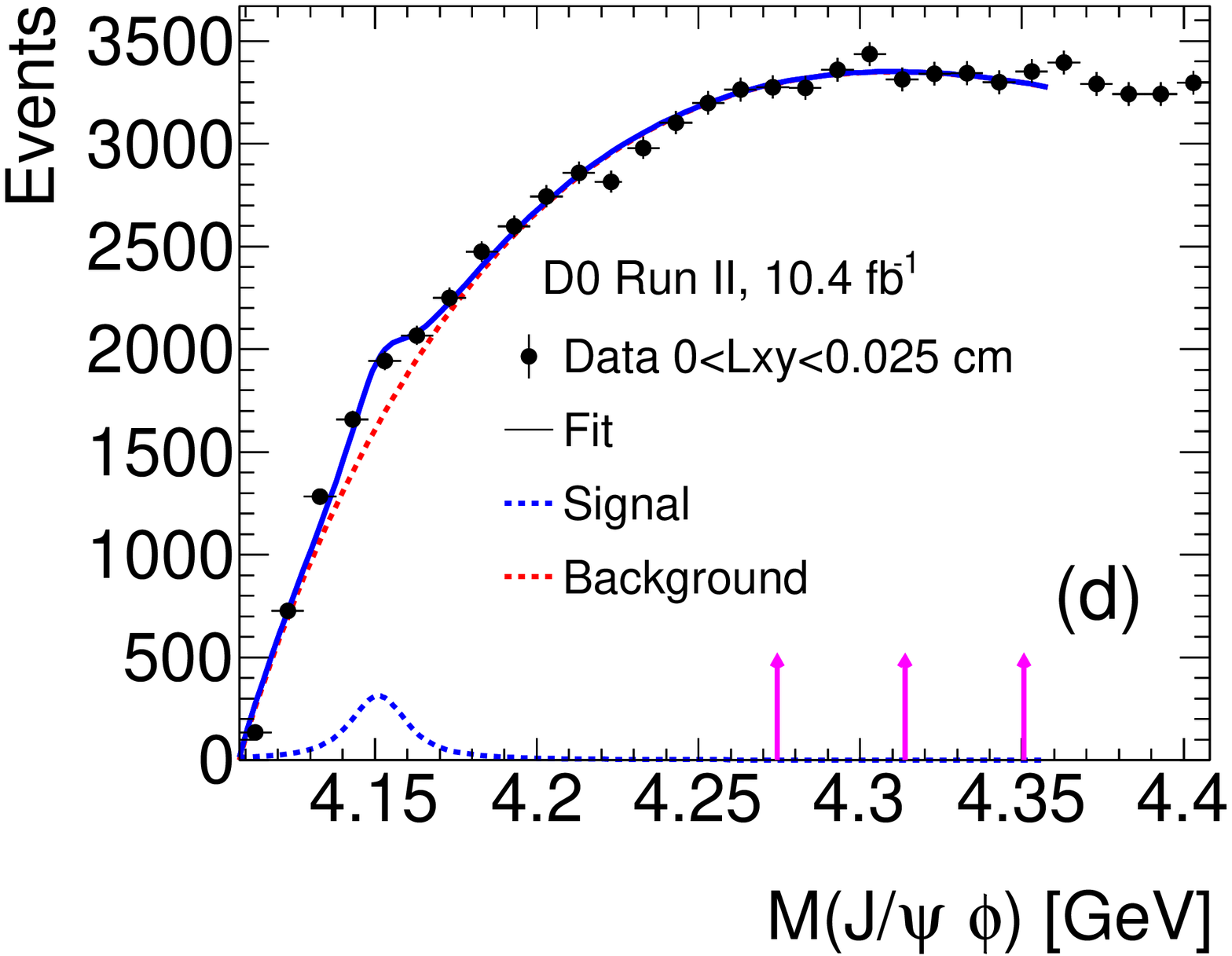} 
  \includegraphics[width=0.8\columnwidth]{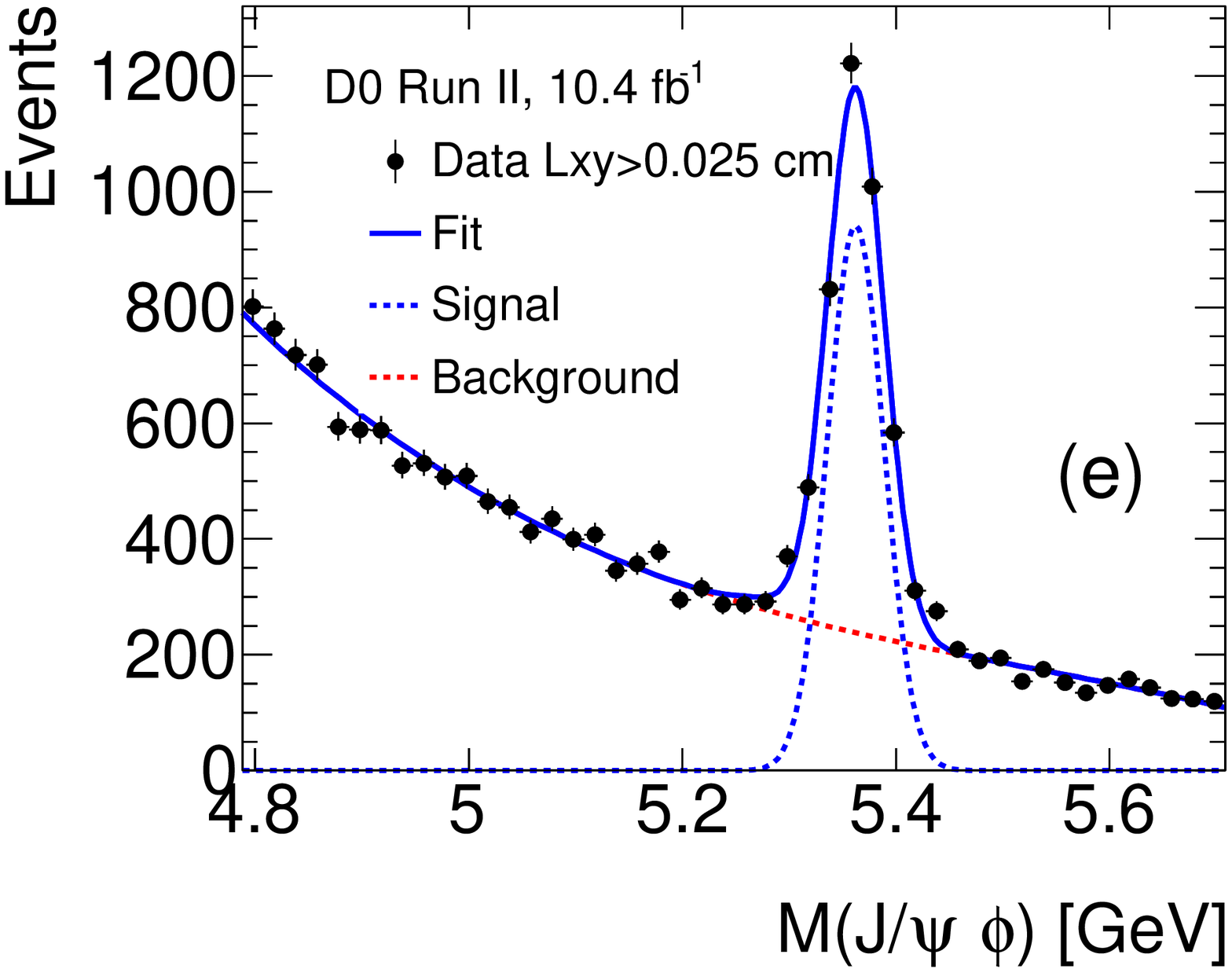}
  \includegraphics[width=0.8\columnwidth]{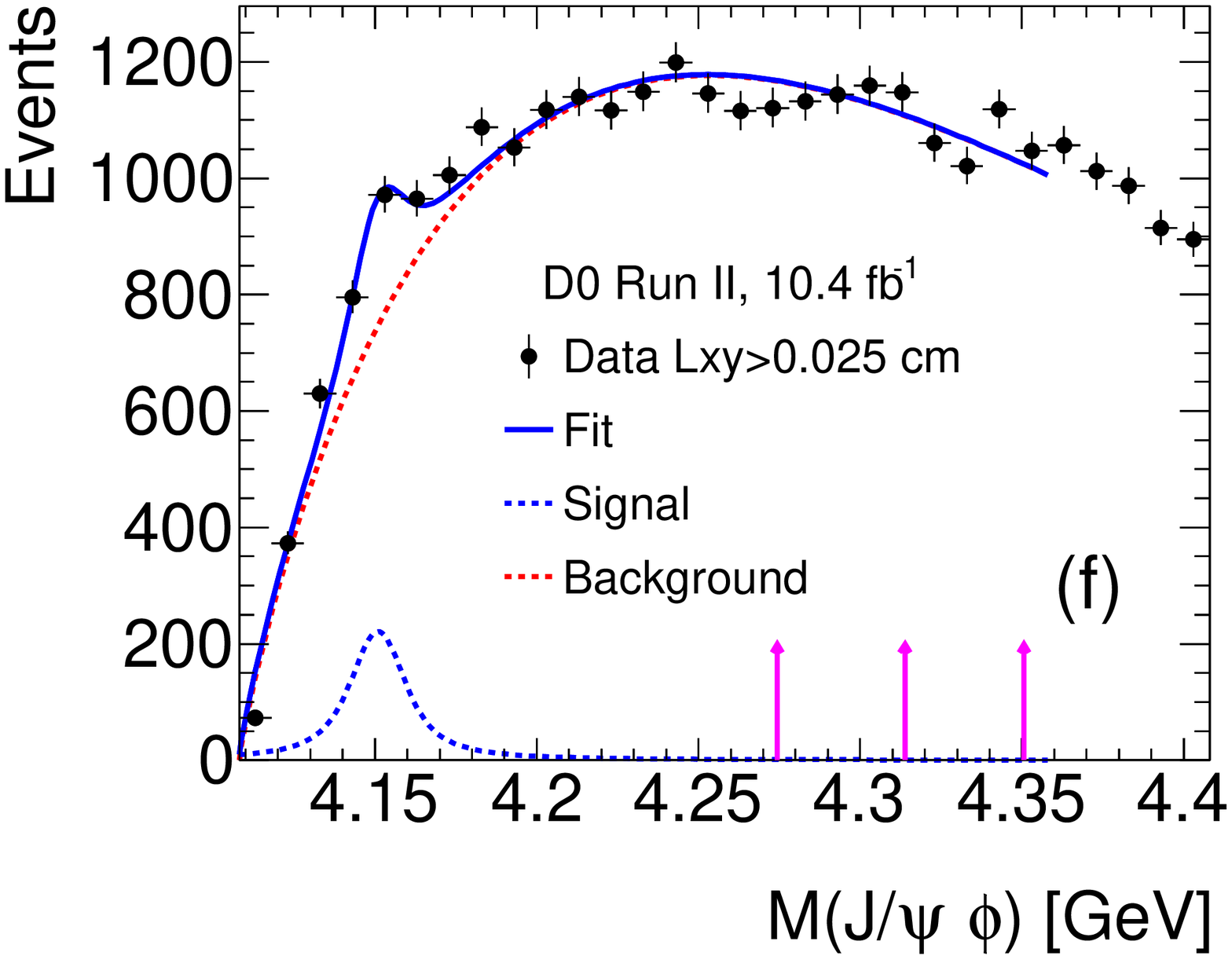} 
 \caption{(color online) Invariant mass distribution of $J/\psi \phi$ candidates in the mass window around 
(left) $B_s^0$
and (right)   $X(4140)$, for events with (a,b)  $-0.025<L_{xy}<0$ cm,
(c,d)  $0<L_{xy}<0.025$ cm and (e,f) $L_{xy}>0.025$  cm.
The arrows indicate the  structures seen by CDF~\cite{Aaltonen:2009at}, CMS~\cite{cms}, 
 and Belle~\cite{Shen:2009vs}.
The signal and background models are described in the text.
}
   \label{fig:mbandx}
\end{figure*}

The mass distributions with superimposed fits for both mass regions and for all three $L_{xy}$
are shown in Fig.~\ref{fig:mbandx}.
The $X(4140)$ and $B_s^0$   yields are presented in Table~\ref{tab:results}.
We also show the expected number of  $X(4140)$ events originating from $b$-hadron decays
in the two low  $L_{xy}$ regions
assuming that the  $L_{xy}$ distribution of the ``non-prompt'' $X(4140)$  is similar
to that of $B_s^0$. For the  Regions 1 and 2,  
we find an excess of signal events, indicating prompt production of $X(4140)$.
For events in Region 2, the increase in the likelihood between
the fit with a free signal yield and the fit with the expected non-prompt contribution only,
 $-2\Delta \ln{\cal L}=23.6$,
corresponds to a statistical significance of 4.9$\sigma$ for  the net prompt signal. The statistical significance of
 the total signal in this  $L_{xy}$ region  is 6.2$\sigma$.
For Region 1, the corresponding values of statistical significance are
3.9$\sigma$ and 4.2$\sigma$.
 If the mass and width parameters are allowed to vary, the fit
for  Region 2 gives the total yield  $N=932\pm216$, $M=4146.8\pm2.4$ MeV, and
 $\Gamma = 15.8 \pm3.8$ MeV.
The data in  Region 1  do not yield a stable fit. Fixing 
the $X(4140)$ mass to 4152.5 MeV in this region, as obtained  in Region 3, we fit a
total yield of   $N=601\pm205$ and  $\Gamma$ of  $19.8 \pm 5.9$~MeV.

 \begin{table*}[htbp]
\caption{Summary of event yields in three $L_{xy}$ regions and their sum for $B_s^0$ and $X(4140)$.
For Regions 1 and 2 the mass of $X(4140)$ is assumed to be 4152.5 MeV and the width is taken to be 16.3 MeV. 
 Also  shown are the  deduced yields for the non-prompt and prompt production of $X(4140)$. The uncertainties are
 statistical.}
\begin{tabular}{ccccc}\hline \hline 
 Parent             & $~~-0.025<L_{xy}<0$ cm ~~   &  $~~0<L_{xy}<0.025$ cm~~ &  $~~L_{xy}>0.025$ cm  &~~  Sum \\\hline
  $B_s^0$           & $191\pm143$   & $804\pm169$ & $3166\pm81$ & $4161\pm236$  \\
  $X(4140)$         & $511\pm120$   & $837\pm135$ & $616\pm 170$ & $1964\pm 248$ \\ 
$X(4140)$ non-prompt& $37\pm26$     & $156\pm 54$ & $616 \pm 170$& $809\pm 175$\\ 
 $X(4140)$ prompt   & $474\pm 123$  & $681\pm149$ &  $\equiv 0$ &  $1155\pm 193$ \\\hline \hline
 \end{tabular}
\label{tab:results}
\end{table*}

 \begin{table*}[htb]
\caption{Summary of systematic uncertainties.}
\begin{tabular}{ccccccc|}\hline
Source                & Mass (MeV)   & Width (MeV) & Rate non-prompt (\%) & Rate prompt (\%)\\\hline
Mass  resolution      &$\pm 0.1$     & $\pm 0.2$   & $\pm 1$   & $\pm 1$   \\    
Mass bias             & $^{+3}_{-0}$      &  --          & --         & --         \\  
Efficiency            & $\pm 4$      &  $\pm 5$  & $\pm 4$ &  $\pm 4$  \\  
Signal model          & $\pm 1$      &  $\pm 2.7$  & $\pm 13$ &  $\pm 15$  \\ 
Fitting range         & $\pm 3$      & $\pm 7.0$   & $\pm 20$  & $\pm 6$ \\
Bin size              & $\pm 1.6$     & $\pm 7.0$ & $\pm 25$  & $\pm 10$ \\
Trigger bias          & --            & --          &  --   & $\pm 5$   \\ 
Mean lifetime         & --            & --          &  $-1.5$   & $+1.5$   \\ \hline
Total                 & $^{+6.2}_{-5.4}$  &  $\pm 11.4$  & $\pm 35$  &  $\pm 19$    \\ \hline \hline
 \end{tabular}
\label{tab:syst}
\end{table*}

There are several uncertainties that may affect measurements of the $X(4140)$ yield,
mass, and width,  the ratio $R$ of the yields of non-prompt $X(4140)$ and $B_s^0$,
and the fraction $f_b$ of all $X(4140)$ events that originate from weak decays of $b$ hadrons.
  
The mass resolution of $4.0$~MeV, obtained in simulations, is in agreement with an approximately linear rise with the released kinetic energy  for decays with a similar topology:  $\psi(2S) \rightarrow J/\psi \pi^+ \pi^-$,  $X(3872) \rightarrow J/\psi \pi^+ \pi^-$,  and the decay $B_s^0  \rightarrow J/\psi \phi$. 
We assign an uncertainty of $\pm 0.1$~MeV to the  resolution at the $X(4140)$ mass.

We   assign an asymmetric uncertainty of 3 MeV  to the $J/\psi \phi$ mass scale
 in the vicinity of $X(4140)$,  based on  the range of the mass deficit between 1 MeV and
5 MeV compared to world-average values,   found  in several channels with this topology.
 We assign the uncertainty in the signal model, taken from the range of
 results obtained with  relativistic and nonrelativistic Breit-Wigner shapes and a relativistic $P$-wave
 Breit-Wigner shape.  
Simulations show  the event reconstruction and selection
efficiency to be independent of the $M(J/\psi \phi)$ invariant  mass, with a possible 
  variation  of $\pm$10\%~\cite{d0x}. The possible variation of the efficiency within the $X(4140)$ mass range affects the
 mass, width, and yield of the signal.
To assess the effects of the fitting procedure and background size and shape, we  vary the fitting mass range
 and bin size.
Some of the single-muon triggers include a trigger term  requiring a presence of tracks
 with non-zero impact parameter. Events recorded solely by such triggers constitute approximately 5\% of all events. Assuming that such triggers are 100\% efficient for events originating from weak decays of $b$ hadrons and
reject all prompt events, we apply a 5\% correction to the prompt yield.
We assign a systematic uncertainty of $\pm5$\% on the fraction $f_b$ due to this correction.
Finally, our assumption of the equality of the relative rates in regions (1) -- (3) for the non-prompt $X(4140)$
 and $B_s^0$ is
based on expectation of the equality of the average  lifetime of  $b$-hadron parents of the $X(4140)$
and that of the $B_s^0$ in the  $J/\psi \phi$ channel. The world-average of the
$B_s^0$  lifetime is 6\% lower than the  lifetime averaged over all $b$ hadron species~\cite{pdg2014}. 
We assign an asymmetric uncertainty in the ratio $R$ and the fraction $f_b$ based on this difference.  
The systematic uncertainties are summarized in Table~\ref{tab:syst}.

We test the stability of the results to the event selection by changing the $\phi$
mass window to $1.012<M(K^+K^-)<1.029$~GeV.
As additional cross-checks, we
perform  fits to subsamples corresponding to the transverse momentum ranges
$5<p_T<10$ GeV and $10<p_T<20$ GeV; to early and late data-taking periods; and to events
in the  central ($|y|<1$) and forward rapidity regions.
In each case, the background shape in the two subsamples
is well described by the same functional form although it requires  different values of the  parameters. In
 all cases  the sums of the resulting signal yields agree with the total yield within a few  events.

Our measured values of the  mass and width of the $X(4140)$ state 
are compared with earlier measurements in Table~\ref{tab:world}. 
The ratio of the $X(4140)$ to $B_s^0$ yield for events with $L_{xy}>0.025$~cm
is $R=0.19 \pm 0.05  {\rm \thinspace (stat)} \pm 0.07  {\rm \thinspace (syst)}$.
After correcting for the efficiency of this $L_{xy}$ cut and for the trigger bias,
 we find the fraction of
$X(4140)$ events originating from $b$ hadrons 
to be $f_b=0.39\pm 0.07  {\rm \thinspace (stat)} \pm 0.10 {\rm \thinspace (syst)} $. 
The yield for the $X(4140)$ state at $L_{xy}>0.025$ cm can also be compared with
the yield of $52\pm19$ events of $X(4140)$ from the decay process $B^+ \rightarrow J/\psi \phi K^+$
  obtained by D0~\cite{d0x} for the same data set.  After correcting 
for a factor of $2.5\pm0.5$  for the efficiency of the full reconstruction of the $B^+$
decay and lower kaon $p_T$ threshold, we expect the yield from the $B^+$ decay
 to be $\approx 130 \pm 60$ events in this analysis.
Our observed yield of $616\pm170$ events exceeds this estimate 
suggesting that decays of $b$ hadrons other than $B^+$  contribute to the non-prompt production of $X(4140)$.

The  $J/\psi \phi$ invariant mass distributions presented in Fig.~\ref{fig:mbandx}  
show no evidence for states in the mass region $4250<M(J/\psi \phi)<4375$ MeV.
Fits allowing for the the states reported by CDF,  CMS, and   Belle, at $L_{xy}>-0.025$ cm,
yield $267\pm276$,  $-283\pm468$, and   $-325\pm254$  events, respectively.
Using the $CL_s$ method~\cite{cls}, we obtain the 95\% upper limits of 744, 557, and 338 events
for the three states. In this upper limit calculation we did not account for systematic uncertainties
 as they were checked to have a negligible impact.
The corresponding 95\% upper limits on the rates relative to the total yield of the $X(4140)$ state
are 0.43, 0.31, and  0.19. 


 \begin{table*}[htb]
\caption{Summary of $X(4140)$ measurements. }
\begin{tabular}{ccccc}\hline \hline 
 Experiment  & Process   &   Mass (MeV)   & Width (MeV) \\\hline
   CDF~\cite{Aaltonen:2009at}   & $B^+ \rightarrow J/\psi \phi K^+$     & $4143.0\pm 2.9 \pm 1.2 $ &  $11.7^{+8.3}_{-5.0} \pm 3.7$\\ 
 CMS~\cite{cms}   & $B^+ \rightarrow J/\psi \phi K^+$     & $4148.0\pm 2.4 \pm 6.3 $ &  $28^{+15}_{-11} \pm 19$\\
  D0~\cite{d0x}   & $B^+ \rightarrow J/\psi \phi K^+$     &  $4159.0\pm4.3  \pm 6.6 $ & $19.9\pm12.6 ^{+3.0}_{-8.0}$ \\
  D0 (this work)   & $\overline p p \rightarrow J/\psi \phi + {\rm anything}$      & $4152.5\pm 1.7 ^{+6.2}_{-5.4} $ &  $16.3 \pm 5.6 \pm 11.4$\\ \hline \hline
 \end{tabular}
\label{tab:world}
\end{table*}

In summary, we have carried out the first search for inclusive production of the state $X(4140)$
in hadronic collisions. We find strong evidence for its direct, prompt production, and observe its production
in weak decays of $b$ hadrons with a rate exceeding the expected rate for the known
decay $B^+ \rightarrow J/\psi \phi K^+$. The significance of the prompt production, including
systematic uncertainties, is 4.7$\sigma$. This is the first evidence for the prompt production of $X(4140)$.
The significance of the non-prompt production, including systematic uncertainties, is 5.6$\sigma$.
The non-prompt production rate of $X(4140)$ relative to
$B_s^0$ observed in the same final state is  $R=0.19 \pm 0.05 {\rm \thinspace (stat)} \pm 0.07 {\rm \thinspace (syst)}$.   
Assuming a relativistic Breit-Wigner line shape, we measure the  mass and width of the $X(4140)$ state to
 be
 $M=4152.5 \pm 1.7 (\rm {stat}) ^{+6.2}_{-5.4} {\rm \thinspace (syst)}$~MeV and width
 $\Gamma=16.3 \pm 5.6 {\rm \thinspace (stat)} \pm 11.4 {\rm \thinspace (syst)}$~MeV, consistent with 
previous measurements~\cite{ Aaltonen:2009at,cms,d0x}.
  
\newpage

We thank the staffs at Fermilab and collaborating institutions,
and acknowledge support from the
Department of Energy and National Science Foundation (United States of America);
Alternative Energies and Atomic Energy Commission and
National Center for Scientific Research/National Institute of Nuclear and Particle Physics  (France);
Ministry of Education and Science of the Russian Federation, 
National Research Center ``Kurchatov Institute" of the Russian Federation, and 
Russian Foundation for Basic Research  (Russia);
National Council for the Development of Science and Technology and
Carlos Chagas Filho Foundation for the Support of Research in the State of Rio de Janeiro (Brazil);
Department of Atomic Energy and Department of Science and Technology (India);
Administrative Department of Science, Technology and Innovation (Colombia);
National Council of Science and Technology (Mexico);
National Research Foundation of Korea (Korea);
Foundation for Fundamental Research on Matter (The Netherlands);
Science and Technology Facilities Council and The Royal Society (United Kingdom);
Ministry of Education, Youth and Sports (Czech Republic);
Bundesministerium f\"{u}r Bildung und Forschung (Federal Ministry of Education and Research) and 
Deutsche Forschungsgemeinschaft (German Research Foundation) (Germany);
Science Foundation Ireland (Ireland);
Swedish Research Council (Sweden);
China Academy of Sciences and National Natural Science Foundation of China (China);
and
Ministry of Education and Science of Ukraine (Ukraine).



\end{document}

%% file: author_list.tex
%
\affiliation{LAFEX, Centro Brasileiro de Pesquisas F\'{i}sicas, Rio de Janeiro, Brazil}
\affiliation{Universidade do Estado do Rio de Janeiro, Rio de Janeiro, Brazil}
\affiliation{Universidade Federal do ABC, Santo Andr\'e, Brazil}
\affiliation{University of Science and Technology of China, Hefei, People's Republic of China}
\affiliation{Universidad de los Andes, Bogot\'a, Colombia}
\affiliation{Charles University, Faculty of Mathematics and Physics, Center for Particle Physics, Prague, Czech Republic}
\affiliation{Czech Technical University in Prague, Prague, Czech Republic}
\affiliation{Institute of Physics, Academy of Sciences of the Czech Republic, Prague, Czech Republic}
\affiliation{Universidad San Francisco de Quito, Quito, Ecuador}
\affiliation{LPC, Universit\'e Blaise Pascal, CNRS/IN2P3, Clermont, France}
\affiliation{LPSC, Universit\'e Joseph Fourier Grenoble 1, CNRS/IN2P3, Institut National Polytechnique de Grenoble, Grenoble, France}
\affiliation{CPPM, Aix-Marseille Universit\'e, CNRS/IN2P3, Marseille, France}
\affiliation{LAL, Universit\'e Paris-Sud, CNRS/IN2P3, Orsay, France}
\affiliation{LPNHE, Universit\'es Paris VI and VII, CNRS/IN2P3, Paris, France}
\affiliation{CEA, Irfu, SPP, Saclay, France}
\affiliation{IPHC, Universit\'e de Strasbourg, CNRS/IN2P3, Strasbourg, France}
\affiliation{IPNL, Universit\'e Lyon 1, CNRS/IN2P3, Villeurbanne, France and Universit\'e de Lyon, Lyon, France}
\affiliation{III. Physikalisches Institut A, RWTH Aachen University, Aachen, Germany}
\affiliation{Physikalisches Institut, Universit\"at Freiburg, Freiburg, Germany}
\affiliation{II. Physikalisches Institut, Georg-August-Universit\"at G\"ottingen, G\"ottingen, Germany}
\affiliation{Institut f\"ur Physik, Universit\"at Mainz, Mainz, Germany}
\affiliation{Ludwig-Maximilians-Universit\"at M\"unchen, M\"unchen, Germany}
\affiliation{Panjab University, Chandigarh, India}
\affiliation{Delhi University, Delhi, India}
\affiliation{Tata Institute of Fundamental Research, Mumbai, India}
\affiliation{University College Dublin, Dublin, Ireland}
\affiliation{Korea Detector Laboratory, Korea University, Seoul, Korea}
\affiliation{CINVESTAV, Mexico City, Mexico}
\affiliation{Nikhef, Science Park, Amsterdam, the Netherlands}
\affiliation{Radboud University Nijmegen, Nijmegen, the Netherlands}
\affiliation{Joint Institute for Nuclear Research, Dubna, Russia}
\affiliation{Institute for Theoretical and Experimental Physics, Moscow, Russia}
\affiliation{Moscow State University, Moscow, Russia}
\affiliation{Institute for High Energy Physics, Protvino, Russia}
\affiliation{Petersburg Nuclear Physics Institute, St. Petersburg, Russia}
\affiliation{Instituci\'{o} Catalana de Recerca i Estudis Avan\c{c}ats (ICREA) and Institut de F\'{i}sica d'Altes Energies (IFAE), Barcelona, Spain}
\affiliation{Uppsala University, Uppsala, Sweden}
\affiliation{Taras Shevchenko National University of Kyiv, Kiev, Ukraine}
\affiliation{Lancaster University, Lancaster LA1 4YB, United Kingdom}
\affiliation{Imperial College London, London SW7 2AZ, United Kingdom}
\affiliation{The University of Manchester, Manchester M13 9PL, United Kingdom}
\affiliation{University of Arizona, Tucson, Arizona 85721, USA}
\affiliation{University of California Riverside, Riverside, California 92521, USA}
\affiliation{Florida State University, Tallahassee, Florida 32306, USA}
\affiliation{Fermi National Accelerator Laboratory, Batavia, Illinois 60510, USA}
\affiliation{University of Illinois at Chicago, Chicago, Illinois 60607, USA}
\affiliation{Northern Illinois University, DeKalb, Illinois 60115, USA}
\affiliation{Northwestern University, Evanston, Illinois 60208, USA}
\affiliation{Indiana University, Bloomington, Indiana 47405, USA}
\affiliation{Purdue University Calumet, Hammond, Indiana 46323, USA}
\affiliation{University of Notre Dame, Notre Dame, Indiana 46556, USA}
\affiliation{Iowa State University, Ames, Iowa 50011, USA}
\affiliation{University of Kansas, Lawrence, Kansas 66045, USA}
\affiliation{Louisiana Tech University, Ruston, Louisiana 71272, USA}
\affiliation{Northeastern University, Boston, Massachusetts 02115, USA}
\affiliation{University of Michigan, Ann Arbor, Michigan 48109, USA}
\affiliation{Michigan State University, East Lansing, Michigan 48824, USA}
\affiliation{University of Mississippi, University, Mississippi 38677, USA}
\affiliation{University of Nebraska, Lincoln, Nebraska 68588, USA}
\affiliation{Rutgers University, Piscataway, New Jersey 08855, USA}
\affiliation{Princeton University, Princeton, New Jersey 08544, USA}
\affiliation{State University of New York, Buffalo, New York 14260, USA}
\affiliation{University of Rochester, Rochester, New York 14627, USA}
\affiliation{State University of New York, Stony Brook, New York 11794, USA}
\affiliation{Brookhaven National Laboratory, Upton, New York 11973, USA}
\affiliation{Langston University, Langston, Oklahoma 73050, USA}
\affiliation{University of Oklahoma, Norman, Oklahoma 73019, USA}
\affiliation{Oklahoma State University, Stillwater, Oklahoma 74078, USA}
\affiliation{Oregon State University, Corvallis, Oregon 97331, USA}
\affiliation{Brown University, Providence, Rhode Island 02912, USA}
\affiliation{University of Texas, Arlington, Texas 76019, USA}
\affiliation{Southern Methodist University, Dallas, Texas 75275, USA}
\affiliation{Rice University, Houston, Texas 77005, USA}
\affiliation{University of Virginia, Charlottesville, Virginia 22904, USA}
\affiliation{University of Washington, Seattle, Washington 98195, USA}
\author{V.M.~Abazov} \affiliation{Joint Institute for Nuclear Research, Dubna, Russia}
\author{B.~Abbott} \affiliation{University of Oklahoma, Norman, Oklahoma 73019, USA}
\author{B.S.~Acharya} \affiliation{Tata Institute of Fundamental Research, Mumbai, India}
\author{M.~Adams} \affiliation{University of Illinois at Chicago, Chicago, Illinois 60607, USA}
\author{T.~Adams} \affiliation{Florida State University, Tallahassee, Florida 32306, USA}
\author{J.P.~Agnew} \affiliation{The University of Manchester, Manchester M13 9PL, United Kingdom}
\author{G.D.~Alexeev} \affiliation{Joint Institute for Nuclear Research, Dubna, Russia}
\author{G.~Alkhazov} \affiliation{Petersburg Nuclear Physics Institute, St. Petersburg, Russia}
\author{A.~Alton$^{a}$} \affiliation{University of Michigan, Ann Arbor, Michigan 48109, USA}
\author{A.~Askew} \affiliation{Florida State University, Tallahassee, Florida 32306, USA}
% ...
%
%
\collaboration{The D0 Collaboration\footnote{with visitors from
$^{a}$Augustana College, Sioux Falls, SD, USA,
$^{b}$The University of Liverpool, Liverpool, UK,
$^{c}$DESY, Hamburg, Germany,
$^{d}$CONACyT, Mexico City, Mexico,
$^{e}$SLAC, Menlo Park, CA, USA,
$^{f}$University College London, London, UK,
$^{g}$Centro de Investigacion en Computacion - IPN, Mexico City, Mexico,
$^{h}$Universidade Estadual Paulista, S\~ao Paulo, Brazil,
$^{i}$Karlsruher Institut f\"ur Technologie (KIT) - Steinbuch Centre for Computing (SCC),
D-76128 Karlsruhe, Germany,
$^{j}$Office of Science, U.S. Department of Energy, Washington, D.C. 20585, USA,
$^{k}$American Association for the Advancement of Science, Washington, D.C. 20005, USA,
$^{l}$Kiev Institute for Nuclear Research, Kiev, Ukraine,
$^{m}$University of Maryland, College Park, Maryland 20742, USA
and
$^{n}$European Orgnaization for Nuclear Research (CERN), Geneva, Switzerland
}} \noaffiliation
\vskip 0.25cm